\documentclass[aps,prd,nofootinbib,amsmath,amssymb,superscriptaddress,twocolumn,10pt]{revtex4}

\pdfoutput=1

\usepackage{graphicx}
\usepackage{dcolumn}
\usepackage{bm}
\usepackage{amssymb}
\usepackage{latexsym}
\usepackage{booktabs}
\usepackage{amsmath}
\usepackage{multirow}
\usepackage[colorlinks=true, linkcolor=red, citecolor=blue]{hyperref}
\usepackage{mathrsfs}
\usepackage{amsfonts}
\usepackage{subfigure}

\newcommand{\ra}[1]{\renewcommand{\arraystretch}{#1}}
\begin{document}

\title{Impacts of dark energy on constraining neutrino mass after Planck 2018}

\author{Ming Zhang}
\affiliation{Department of Physics, College of Sciences, Northeastern University, Shenyang
110819, China}
\author{Jing-Fei Zhang}
\affiliation{Department of Physics, College of Sciences, Northeastern University, Shenyang
110819, China}
\author{Xin Zhang\footnote{Corresponding author}}
\email{zhangxin@mail.neu.edu.cn}
\affiliation{Department of Physics, College of Sciences, Northeastern University, Shenyang
110819, China}
\affiliation{Ministry of Education's Key Laboratory of Data Analytics and Optimization for Smart Industry, Northeastern University, Shenyang 110819, China}
\affiliation{Center for High Energy Physics, Peking University, Beijing 100080, China}
\affiliation{Center for Gravitation and Cosmology, Yangzhou University, Yangzhou 225009, China}

\begin{abstract}
Considering the mass splittings of three active neutrinos, we investigate how the nature of dark energy affects the cosmological constraints on the total neutrino mass $\sum m_\nu$ using the latest cosmological observations. In this paper, some typical dark energy models, including $\Lambda$CDM, $w$CDM, CPL, and HDE models, are discussed. In the analysis, we also consider the effects from the neutrino mass hierarchies, i.e., the degenerate hierarchy (DH), the normal hierarchy (NH), and the inverted hierarchy (IH). We employ the current cosmological observations to do the analysis, including the Planck 2018 temperature and polarization power spectra, the baryon acoustic oscillations (BAO), the type Ia supernovae (SNe), and the Hubble constant $H_0$ measurement. In the $\Lambda$CDM+$\sum m_\nu$ model, we obtain the upper limits of the neutrino mass $\sum m_\nu < 0.123$ eV (DH), $\sum m_\nu < 0.156$ eV (NH), and $\sum m_\nu < 0.185$ eV (IH) at the $95\%$ C.L., using the Planck+BAO+SNe data combination. For the $w$CDM+$\sum m_\nu$ model and the CPL+$\sum m_\nu$ model, larger upper limits of $\sum m_\nu$ are obtained compared to those of the $\Lambda$CDM+$\sum m_\nu$ model. The most stringent constraint on the neutrino mass, $\sum m_\nu<0.080$ eV (DH), is derived in the HDE+$\sum m_\nu$ model. In addition, we find that the inclusion of the local measurement of the Hubble constant in the data combination leads to tighter constraints on the total neutrino mass in all these dark energy models.

\end{abstract}

\maketitle

\section{Introduction}

Neutrino oscillation experiments \cite{Fukuda:1998mi,Ahmad:2002jz} indicate that the three neutrino flavor eigenstates ($\nu_e$, $\nu_\mu$, $\nu_\tau$) are actually quantum superpositions of the three mass eigenstates ($\nu_1$, $\nu_2$, $\nu_3$) with masses $m_1$, $m_2$, and $m_3$. However, neutrino oscillation experiments cannot measure the absolute neutrino masses, but can only give the squared mass differences between the different mass eigenstates of neutrino. The solar and reactor experiments gave the result of $\Delta m_{21}^2 \simeq 7.54\times 10^{-5}$ eV$^{2}$ and the atmospheric and accelerator beam experiments gave the result of $|\Delta m_{31}^2| \simeq 2.46\times 10^{-3}$ eV$^{2}$ \cite{Agashe:2014kda}, which indicates that there are two possible neutrino mass hierarchies, i.e., the normal hierarchy (NH) with $m_{1}<m_{2}\ll m_{3}$ and the inverted hierarchy (IH) with $m_{3}\ll m_{1}< m_{2}$. The case of neglecting the neutrino mass splitting, namely $m_{1} = m_{2} = m_{3}$, is called the degenerate hierarchy (DH).

Nevertheless, cosmological observations could provide a useful tool to measure the absolute neutrino total mass. With the decrease of the neutrino temperature, neutrino becomes non-relativistic at $T \sim 0.15$ eV in the evolution of the universe. Then the mass effect of neutrinos begins to appear, which leads to a nonnegligible influence on the cosmic microwave background (CMB) and large-scale structure (LSS) \cite{Abazajian:2013oma,Lesgourgues:2006nd,Wong:2011ip,Lesgourgues:2012uu,Lesgourgues:2014zoa,Archidiacono:2016lnv,Lattanzi:2017ubx}.
Therefore, we could extract much useful information about neutrino from cosmological observations.

Recently, the observational data of Planck 2018 have been released by the Planck collaboration, and according to the latest data the limit of the total neutrino mass is $\sum m_\nu < 0.24$ eV (95\% C.L., TT,TE,EE+lowE+lensing) \cite{Aghanim:2018eyx}. Since the baryon acoustic oscillations (BAO) data at low redshifts can break the geometric degeneracy inherent in CMB, the combination of the acoustic scales measured by the CMB and BAO data can determine the background geometry sufficiently. Combining BAO data with CMB data, the neutrino mass can be constrained to be $\sum m_\nu < 0.12$ eV (95\% C.L., TT,TE,EE+lowE+lensing+BAO). Adding the Pantheon type Ia supernovae (SNe) luminosity distance measurements, the constraint only becomes slightly better, with the result still roughly $\sum m_\nu < 0.11$ eV (95\% C.L., TT,TE,EE+lowE+lensing+BAO+SNe). It is noted that these results are based on the $\Lambda$CDM+$\sum m_\nu$ model.

Therefore, we wish to investigate the impacts of dark energy on constraining the total neutrino mass. In this work, we consider some typical dark energy models, including the $\Lambda$CDM model, the $w$CDM model, the $w_0w_a$CDM model (also known as the
Chevallier-Polarski-Linder model or the CPL model) \cite{Chevallier:2000qy,Linder:2002et}, and the holographic dark energy (HDE) model \cite{Li:2004rb,Huang:2004ai,Zhang:2014ija,Wang:2016och,Wang:2013zca,Cui:2015oda,He:2016rvp,Xu:2016grp}. In addition, we also consider the effects from the neutrino mass hierarchies (i.e., DH, NH, and IH) in our analysis.

More recently, some related studies of constraints on the total neutrino mass have been made; see, e.g.,  Refs.~\cite{RoyChoudhury:2019hls,Reid:2009nq,Thomas:2009ae,Carbone:2010ik,Li:2012vn,Wang:2012uf,Audren:2012vy,Riemer-Sorensen:2013jsa,Zhang:2014dxk,Zhang:2014nta,Zhang:2014ifa,Palanque-Delabrouille:2014jca,Li:2015poa,Zhang:2015rha,Geng:2015haa,Chen:2015oga,Allison:2015qca,Cuesta:2015iho,Chen:2016eyp,Lu:2016hsd,Kumar:2016zpg,Xu:2016ddc,Vagnozzi:2017ovm,Zhang:2017rbg,Zhang:2015uhk,Lorenz:2017fgo,Zhao:2017jma,Vagnozzi:2018jhn,Wang:2018lun,Zhang:2019ipd,Guo:2018gyo,Zhao:2018fjj,Loureiro:2018pdz,Huang:2015wrx,Bellomo:2016xhl,Hu:2014hra,Xu:2016jns,Wang:2016tsz,Feng:2019mym}. The cosmological constraints on the total neutrino mass in dynamical dark energy models have been discussed in, e.g., Refs.~\cite{Zhang:2017rbg,Zhang:2015uhk}, which indicates that the nature of dark energy can have a significant influence on the measurement of the total neutrino mass. As the latest CMB data have been released by the Planck collaboration, the results need to be updated. In this work, we employ the latest cosmological observations, including the CMB, BAO, SNe, and $H_0$ data to make a new analysis.

In this work, we will use the recent local measurement of the Hubble constant $H_0$, with the result of $H_0 = 74.03\pm1.42$ km s$^{-1}$ Mpc$^{-1}$ (68\% C.L.), by Riess et al. \cite{Riess:2019cxk}. Note that this local measurement result is in more than 4 $\sigma$ tension with the result of the Planck 2018 observation assuming a 6-parameter base-$\Lambda$ CDM model. Thus, we also wish to investigate how the inclusion of the $H_0$ local measurement would affect the measurement of the total neutrino mass in these dark energy models.

This paper is organized as follows. In Sec.{~\ref{sec:2}}, we describe the methodology in our analysis. In Sec.{~\ref{sec:3}}, we show the results and make some discussions. Finally, the conclusion is given in Sec.{~\ref{sec:4}}.

\section{methodology}\label{sec:2}

We take into account the neutrino mass splittings between the three active neutrinos. We employ the measurement results of neutrino oscillation experiments \cite{Agashe:2014kda},
\begin{equation}\label{1}
\Delta m_{21}^{2} \equiv m_{2}^{2} - m_{1}^{2} = 7.54\times 10^{-5} \rm{eV}^{2},
\end{equation}
\begin{equation}\label{2}
|\Delta m_{31}^{2}| \equiv |m_{3}^{2} - m_{1}^{2}| = 2.46\times 10^{-3} \rm{eV}^{2}.
\end{equation}

The total neutrino mass $\sum m_\nu$ is the sum of three active neutrino mass. For the NH case, $\sum m_\nu$ is written as
\begin{equation}\label{3}
\sum m_\nu^{\rm{NH}} = m_{1} + \sqrt{m_{1}^{2} + \Delta m_{21}^{2}} + \sqrt{m_{1}^{2} + \Delta m_{31}^{2}},
\end{equation}
where $m_{1}$ is a free parameter. For the IH case, $\sum m_\nu$ is written as
\begin{equation}\label{4}
\sum m_\nu^{\rm{IH}} = \sqrt{m_{3}^{2} + |\Delta m_{31}^{2}|} + \sqrt{m_{3}^{2} + |\Delta m_{31}^{2}| + \Delta m_{21}^{2}} + m_{3},
\end{equation}
where $m_{3}$ is a free parameter. For the DH case, ignoring the neutrino mass splittings, $\sum m_\nu$ can be written as
\begin{equation}\label{5}
\sum m_\nu^{\rm{DH}} = m_{1} + m_{2} + m_{3} = 3 m,
\end{equation}
where $m$ is a free parameter. Therefore, the lower bounds of $\sum m_\nu$ are $0$ eV, $0.06$ eV and $0.1$ eV for DH, NH and IH, respectively. In this way, the total neutrino mass $\sum m_\nu$ as an additional parameter will be considered in our analysis.

In this paper, we make a global fit analysis on the different dark energy models, i.e., the $\Lambda$CDM+$\sum m_\nu$ model, the $w$CDM+$\sum m_\nu$ model, the CPL+$\sum m_\nu$ model, and the HDE+$\sum m_\nu$ model. We modify the publicly available Markov chain Monte-Carlo package {\tt CosmoMC} \cite{Lewis:2002ah} (that uses the Boltzmann solver {\tt CAMB} \cite{Lewis:1999bs}) to do the numerical calculations.

\begin{table*}[htb]
\caption{Fitting results of the cosmological parameters in the $\Lambda$CDM+$\sum m_\nu$ model for three neutrino mass hierarchy cases, i.e., the DH case, the NH case, and the IH case, using the Planck+BAO+SNe and Planck+BAO+SNe+$H_0$ data combinations.}\label{table1}
\centering
\ra{1.3}
\resizebox{\textwidth}{!}{\begin{tabular}{@{}rcrrrcrrr@{}}\toprule
\hline
\multicolumn{1}{c}{Data} & \phantom{abc} & \multicolumn{3}{c}{Planck+BAO+SNe} & \phantom{abc} & \multicolumn{3}{c}{Planck+BAO+SNe+$H_0$}\\
\cmidrule{1-1} \cmidrule{3-5} \cmidrule{7-9}
\cline{1-1}  \cline{3-5} \cline{7-9}
\multicolumn{1}{c}{Mass ordering} && DH & NH & IH & & DH & NH & IH  \\
\hline
%
%
%
%
%
$H_0$ [km s$^{-1}$ Mpc$^{-1}$]   && $67.75\pm0.49$ & $67.48\pm0.47$ & $67.26\pm0.45$ &
                                  & $68.40\pm0.44$ & $68.11\pm0.43$ & $67.88\pm0.43$ \\

$\Omega_{\rm m}$       && $0.3097\pm0.0063$ & $0.3126\pm0.0063$ & $0.3150\pm0.0060$ &
                        & $0.3015\pm0.0056$ & $0.3044\pm0.0056$ & $0.3069\pm0.0056$ \\

$\sigma_8$            && $0.812^{+0.013}_{-0.008}$ & $0.801^{+0.011}_{-0.008}$ & $0.793^{+0.010}_{-0.008}$ &
                       & $0.813^{+0.010}_{-0.008}$ & $0.801^{+0.009}_{-0.008}$ & $0.792^{+0.009}_{-0.008}$ \\

$\sum m_\nu$ [eV]      && $<0.123$  & $<0.156$ & $<0.185$ &
                        & $<0.082$ & $<0.125$ & $<0.160$   \\
\hline
$\chi^{2}$             &&$3805.133$  &$3807.205$  &$3809.012$  &
                        &$3821.466$  &$3825.557$  &$3828.810$  \\
\hline
\end{tabular}}
\end{table*}

\begin{figure*}[htbp]
\begin{center}
\includegraphics[width=8.5cm]{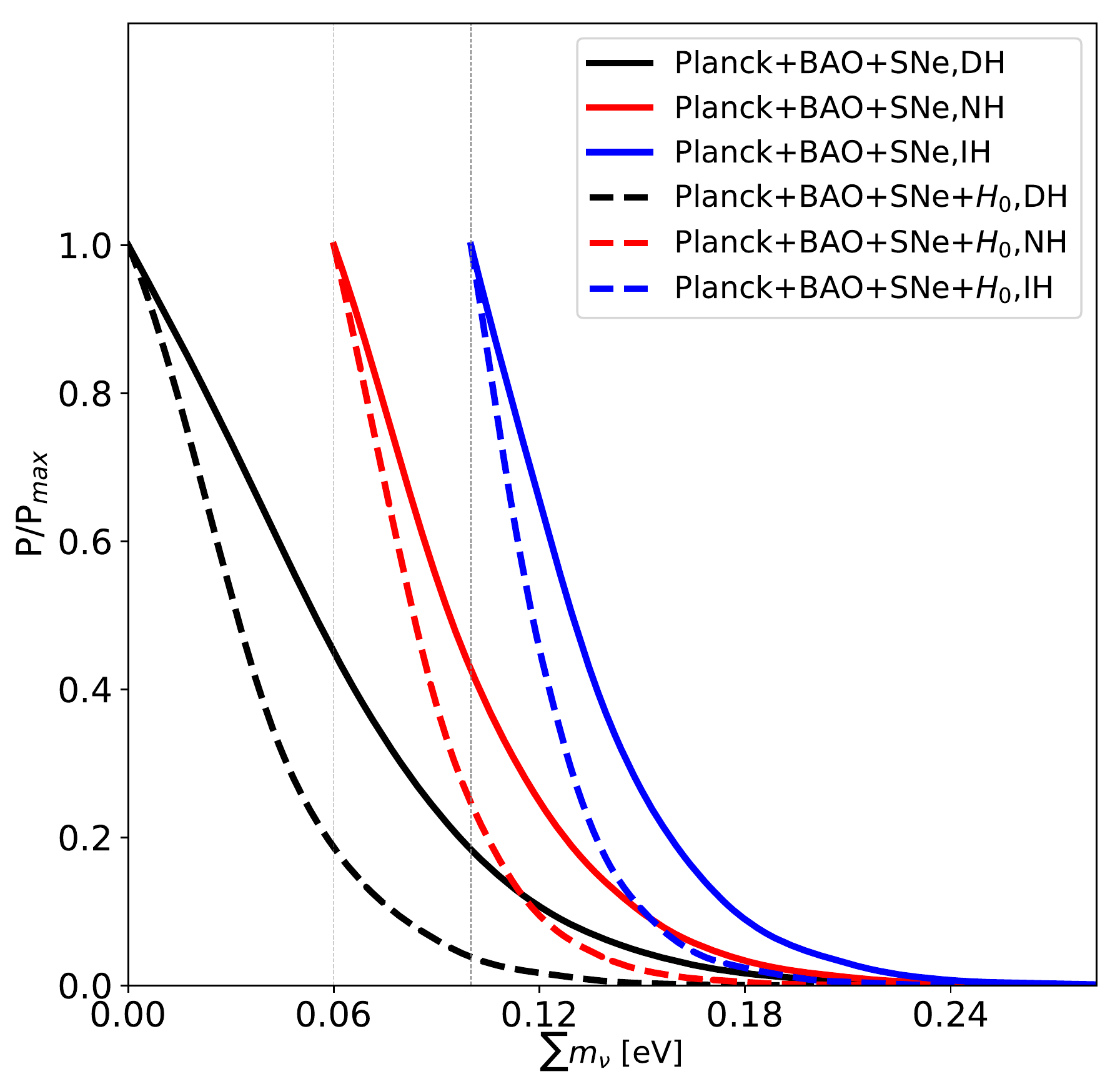}
\includegraphics[width=8.5cm]{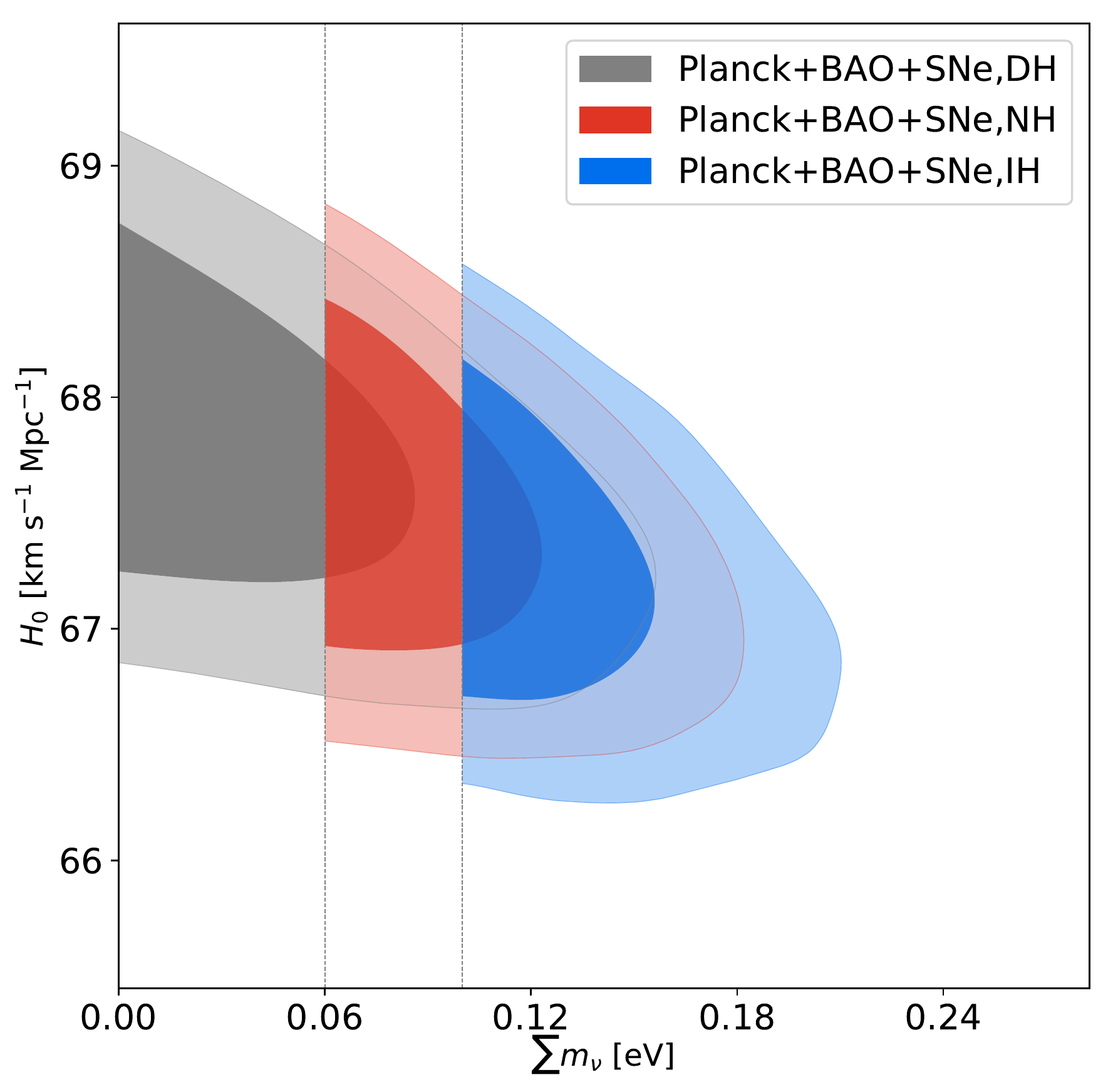}
\end{center}
\caption{$Left$: The one-dimensional marginalized posterior distributions for $\sum m_\nu$ using the Planck+BAO+SNe and Planck+BAO+SNe+$H_0$ data combinations in the $\Lambda$CDM+$\sum m_\nu$ model. $Right$: The two-dimensional marginalized contours ($1 \sigma$ and $2 \sigma$) in the $\sum m_\nu$-$H_0$ plane for three neutrino mass hierarchy cases, i.e., the DH case, the NH case, and the IH case, by using Planck+BAO+SNe data combination in the $\Lambda$CDM+$\sum m_\nu$ model.}
\label{figure1}
\end{figure*}

Here, we give a brief introduction to these dark energy models.

\begin{itemize}

\item \textbf{The $\Lambda$CDM+$\sum m_\nu$ model:} The model containing a cosmological constant $\Lambda$ and cold dark matter is called the $\Lambda$CDM model, which can fit various cosmological observations well. For the $\Lambda$CDM+$\sum m_\nu$ model, the parameter space vector is:
\begin{equation}
P_1 \equiv \left(\omega_b, \omega_c, \Theta_{\rm{s}}, \tau, n_{\rm{s}}, \rm{ln}[10^{10} A_{\rm{s}}], \sum m_\nu \right),
\end{equation}
where $\omega_b \equiv \Omega_b h^2$ and $\omega_c \equiv \Omega_c h^2$ represent baryon and cold dark matter densities, respectively, $\Theta_{\rm{s}}$ is the ratio between sound horizon $r_s$ and angular diameter distance $D_{\rm{A}}$ at the time of photon decoupling, $\tau$ is the optical depth to the reionization of the universe, $n_{\rm{s}}$ and $A_{\rm{s}}$ are the power-law spectral index and amplitude of the power spectrum of primordial curvature perturbations, respectively, and $\sum m_\nu$ is the total neutrino mass.
\item \textbf{The $w$CDM+$\sum m_\nu$ model:} The $w$CDM model is the simplest dynamical dark energy model, in which the equation-of-state (EoS) parameter $w(z)$ is assumed to be a constant. For the $w$CDM+$\sum m_\nu$ model, the parameter space vector is:
\begin{equation}
P_2 \equiv \left(\omega_b, \omega_c, \Theta_{\rm{s}}, \tau, n_{\rm{s}}, \rm{ln}[10^{10} A_{\rm{s}}], \emph{w}, \sum m_\nu \right).
\end{equation}
\item \textbf{The CPL+$\sum m_\nu$ model:} For probing the evolution of $w(z)$, the most widely used parametrization model is the CPL model (also called the $w_0 w_a$CDM model) \cite{Chevallier:2000qy,Linder:2002et}. The form of $w(z)$ in this model is given by
\begin{equation}
w(z) = w_0 + w_a (1-a) = w_0 + w_a \frac{z}{1+z},
\end{equation}
where $w_0$ and $w_a$ are free parameters. So, for the CPL+$\sum m_\nu$ model, the parameter space vector is:
\begin{equation}
P_3 \equiv \left(\omega_b, \omega_c, \Theta_{\rm{s}}, \tau, n_{\rm{s}}, \rm{ln}[10^{10} A_{\rm{s}}], \emph{w}_0, \emph{w}_\emph{a}, \sum m_\nu \right).
\end{equation}
\item \textbf{The HDE+$\sum m_\nu$ model:} The HDE model is built based on the the effective quantum field theory together with the holographic principle of quantum gravity. We can put an energy bound on the vacuum energy density, $\rho_{\rm {de}} L^{3} \leq M^2_{\rm {Pl}} L$, where $M_{\rm {Pl}}$ is the reduced Planck mass, which means that the total energy in a spatial region with size $L$ should not exceed the mass of a black hole with the same size \cite{Cohen:1998zx}. The largest length size that is compatible with this bound is the infrared cutoff size of this effective quantum field theory. An infrared scale can saturate that bound, and thus the dark energy density can be written as \cite{Li:2004rb}
\begin{equation}
\rho_{\rm {de}} = 3 c^2 M^2_{\rm {Pl}} L^{-2},
\end{equation}
where $c$ is a dimensionless phenomenological parameter (note that here c is not the speed of light), which plays an important role in determining the properties of the holographic dark energy. The value of $c$ determines the evolution of $w$. In the HDE model, the EoS can be expressed as
\begin{equation}
w = - \frac{1}{3} - \frac{2}{3} \frac{\sqrt{\Omega_{\rm{de}}}}{c}.
\end{equation}
According to this equation, we can find that in the early times $w \rightarrow -1/3$ (since $\Omega_{\rm{de}} \rightarrow 0$) and in the far future $w \rightarrow -1/3 - 2/(3c)$ (since $\Omega_{\rm{de}} \rightarrow 1$). Thus, when $c < 1$, we can find that the EoS parameter $w$ crosses $-1$ during the cosmological evolution. For the HDE+$\sum m_\nu$ model, the parameter space vector is:
\begin{equation}
P_4 \equiv \left(\omega_b, \omega_c, \Theta_{\rm{s}}, \tau, n_{\rm{s}}, \rm{ln}[10^{10} A_{\rm{s}}], \emph{c}, \sum m_\nu \right).
\end{equation}
\end{itemize}

\begin{table*}[tp]
\caption{Fitting results of the cosmological parameters in the $w$CDM+$\sum m_\nu$ model for three neutrino mass hierarchy cases, i.e., the DH case, the NH case, and the IH case, using the Planck+BAO+SNe and Planck+BAO+SNe+$H_0$ data combinations.}\label{table2}
\centering
\ra{1.3}
\resizebox{\textwidth}{!}{\begin{tabular}{@{}rcrrrcrrr@{}}\toprule
\hline
 \multicolumn{1}{c}{Data} & \phantom{abc} & \multicolumn{3}{c}{Planck+BAO+SNe} & \phantom{abc} & \multicolumn{3}{c}{Planck+BAO+SNe+$H_0$}\\
\cmidrule{1-1} \cmidrule{3-5} \cmidrule{7-9}
\cline{1-1} \cline{3-5} \cline{7-9}
\multicolumn{1}{c}{Mass ordering} && DH & NH & IH && DH & NH & IH\\
\hline
%
%
%
%
%
$w$     && $-1.029\pm0.035$ & $-1.042\pm0.035$ & $-1.051\pm0.035$ &
         & $-1.078\pm0.033$ & $-1.090\pm0.033$ & $-1.100^{+0.034}_{-0.031}$ \\

$H_0$ [km s$^{-1}$ Mpc$^{-1}$]           && $68.27\pm0.83$ & $68.23\pm0.83$ & $68.21\pm0.81$ &
                                          & $69.79\pm0.73$ & $69.74\pm0.73$ & $69.70\pm0.74$ \\

$\Omega_{\rm m}$       && $0.3064\pm0.0078$ & $0.3076\pm0.0078$ & $0.3084\pm0.0076$ &
                        & $0.2932\pm0.0066$ & $0.2945\pm0.0066$ & $0.2954\pm0.0067$ \\

$\sigma_8$             && $0.819\pm0.015$ & $0.811^{+0.015}_{-0.014}$ & $0.805\pm0.014$ &
                        & $0.834^{+0.015}_{-0.013}$ & $0.826^{+0.014}_{-0.013}$ & $0.820\pm0.014$ \\

$\sum m_\nu$ [eV]      && $<0.155$  & $<0.195$ & $<0.220$ &
                        & $<0.145$ & $<0.183$ & $<0.210$   \\
\hline
$\chi^{2}$             && $3805.053$  &$3806.381$  &$3807.724$  &
                        & $3817.072$  &$3818.757$  &$3819.912$  \\
\hline
\end{tabular}}
\end{table*}

\begin{figure*}[htbp]
\begin{center}
\includegraphics[width=8.5cm]{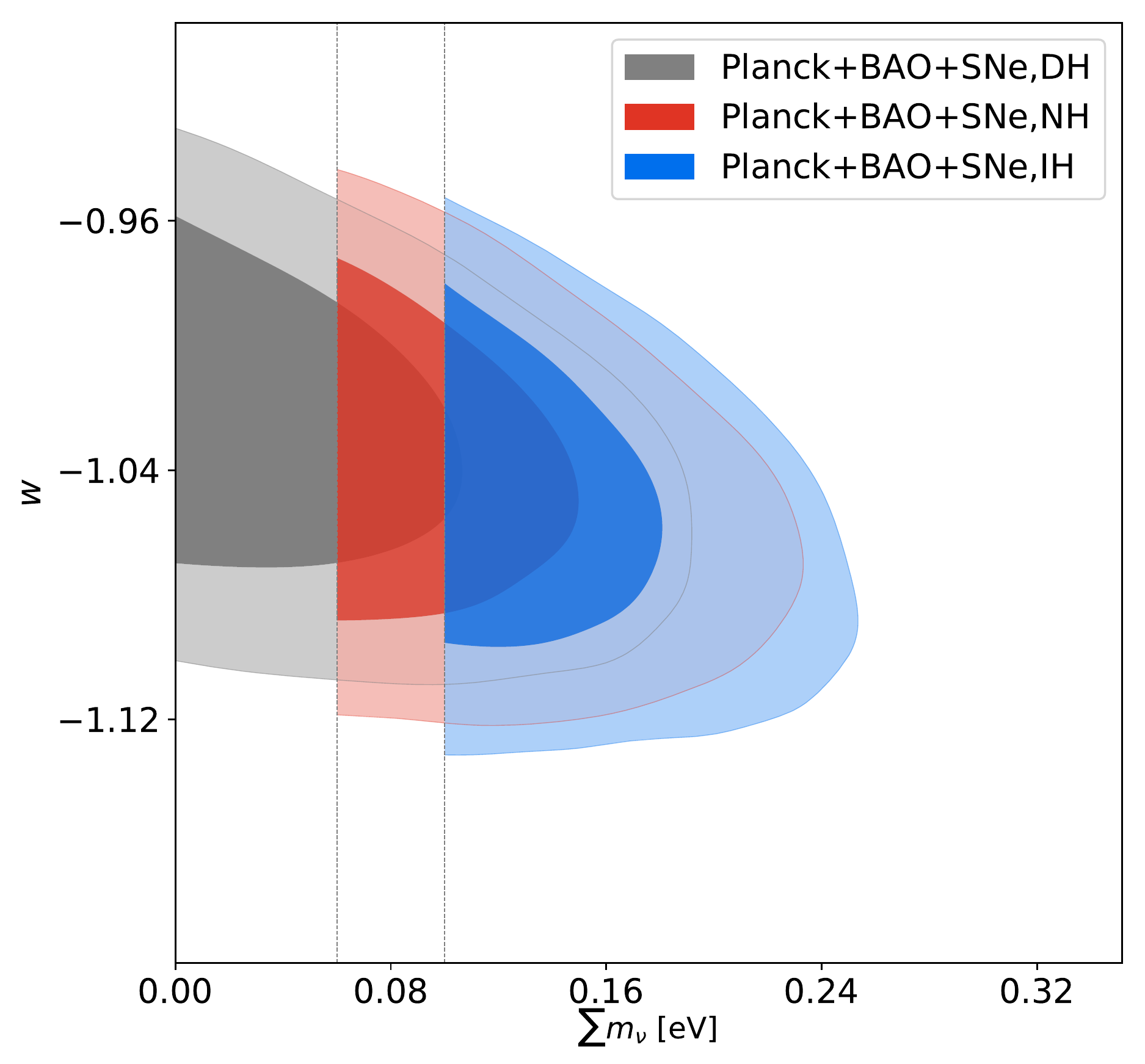}
\includegraphics[width=8.5cm]{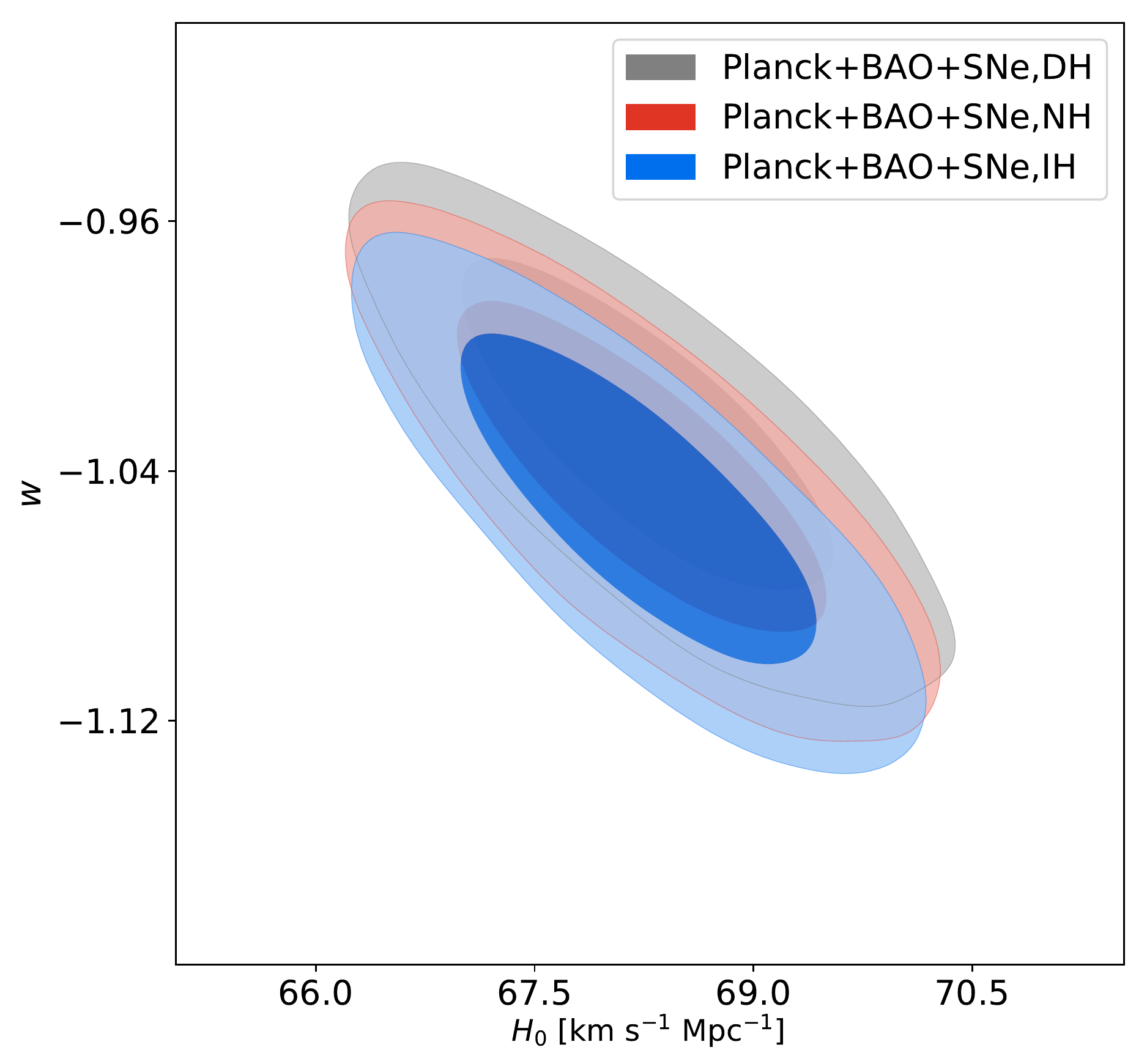}
\end{center}
\caption{The two-dimensional marginalized contours ($1 \sigma$ and $2 \sigma$) in the $\sum m_\nu$-$w$ and $H_0$-$w$ planes for three neutrino mass hierarchy cases, i.e., the DH case, the NH case, and the IH case, by using Planck+BAO+SNe data combination in the $w$CDM+$\sum m_\nu$ model.}
\label{figure2}
\end{figure*}

The observational data sets used in this work include CMB, BAO, SNe, and $H_0$. Here we also briefly describe these observational data.

\begin{itemize}

\item \textbf{The CMB data:} We employ the CMB likelihood including the TT, TE, and EE spectra at $\ell \geq 30$, the low-$\ell$ temperature Commander likelihood, and the low-$\ell$ SimAll EE likelihood, from the Planck 2018 release \cite{Aghanim:2018eyx}.
\item \textbf{The BAO data:} We employ the measurements of the BAO signals from different galaxy surveys, including the DR7 Main Galaxy Sample (MGS) at the effective redshift of $z_{\rm eff}=0.15$ \cite{Ross:2014qpa}, the Six-degree-Field Galaxy Survey (6dFGS) at $z_{\rm eff}=0.106$ \cite{Beutler:2011hx}, and the latest BOSS data release 12 (DR12) in three redshift slices of $z_{\rm eff}=0.38$, $0.51$, and $0.61$ \cite{Alam:2016hwk}.
\item \textbf{The SNe data:} We use the latest SNe data given the Pantheon Sample \cite{Scolnic:2017caz}, which contains 1048 SNe data in the redshift range of $0.01 < z <2.3$.
\item \textbf{The Hubble constant:} We use the result of the direct measurement of the Hubble constant, with the result of $H_{0}=74.03 \pm 1.42$ km s$^{-1}$ Mpc$^{-1}$, given by Riess et al. \cite{Riess:2019cxk}.

\end{itemize}

In this study, our basic data combination is Planck+BAO+SNe. In addition, in order to investigate the impacts of the $H_0$ measurement on constraints on the neutrino mass, we also consider the data combination of Planck+BAO+SNe+$H_0$.

\section{results and discussion}\label{sec:3}

\begin{table*}[htbp]
\caption{Fitting results of the cosmological parameters in the CPL+$\sum m_\nu$ model for three neutrino mass hierarchy cases, i.e., the DH case, the NH case, and the IH case, using the Planck+BAO+SNe and Planck+BAO+SNe+$H_0$ data combinations.}\label{table3}
\centering
\ra{1.3}
\resizebox{\textwidth}{!}{\begin{tabular}{@{}rcrrrcrrr@{}}\toprule
\hline
\multicolumn{1}{c}{Data} & \phantom{abc} & \multicolumn{3}{c}{Planck+BAO+SNe} & \phantom{abc} & \multicolumn{3}{c}{Planck+BAO+SNe+$H_0$}\\
\cmidrule{1-1} \cmidrule{3-5} \cmidrule{7-9}
\cline{1-1} \cline{3-5} \cline{7-9}
\multicolumn{1}{c}{Mass ordering} && DH & NH & IH && DH & NH & IH\\
\hline
%
%
%
%
%
$w_0$   && $-0.945\pm0.087$ & $-0.933\pm0.089$ & $-0.923\pm0.089$ &
         & $-1.003\pm0.082$ & $-0.988\pm0.086$ & $-0.978\pm0.088$ \\

$w_a$   && $-0.41^{+0.44}_{-0.30}$ & $-0.52^{+0.46}_{-0.32}$ & $-0.61^{+0.46}_{-0.33}$ &
         & $-0.38^{+0.41}_{-0.31}$ & $-0.50^{+0.44}_{-0.33}$ & $-0.60^{+0.45}_{-0.34}$ \\

$H_0$ [km s$^{-1}$ Mpc$^{-1}$]    && $68.22\pm0.83$ & $68.19\pm0.83$ & $68.14\pm0.84$ &
                                   & $69.78\pm0.73$ & $69.71\pm0.74$ & $69.69\pm0.73$ \\

$\Omega_{\rm m}$       && $0.3087\pm0.0082$ & $0.3102\pm0.0083$ & $0.3113\pm0.0083$ &
                        & $0.2948\pm0.0068$ & $0.2965\pm0.0070$ & $0.2976\pm0.0069$ \\

$\sigma_8$             && $0.819^{+0.018}_{-0.015}$ & $0.813^{+0.018}_{-0.015}$ & $0.808^{+0.018}_{-0.015}$ &
                        & $0.835^{+0.017}_{-0.015}$ & $0.828^{+0.017}_{-0.014}$ & $0.823^{+0.017}_{-0.014}$ \\

$\sum m_\nu$ [eV]      && $<0.247$ & $<0.290$ & $<0.305$ &
                        & $<0.216$ & $<0.255$ & $<0.281$   \\
\hline
$\chi^{2}$             && $3804.644$  &$3805.938$  &$3806.531$  &
                        & $3816.716$  &$3817.806$  &$3818.809$  \\
\hline
\end{tabular}}
\end{table*}

\begin{figure*}[htbp]
\begin{center}
\includegraphics[width=5.8cm]{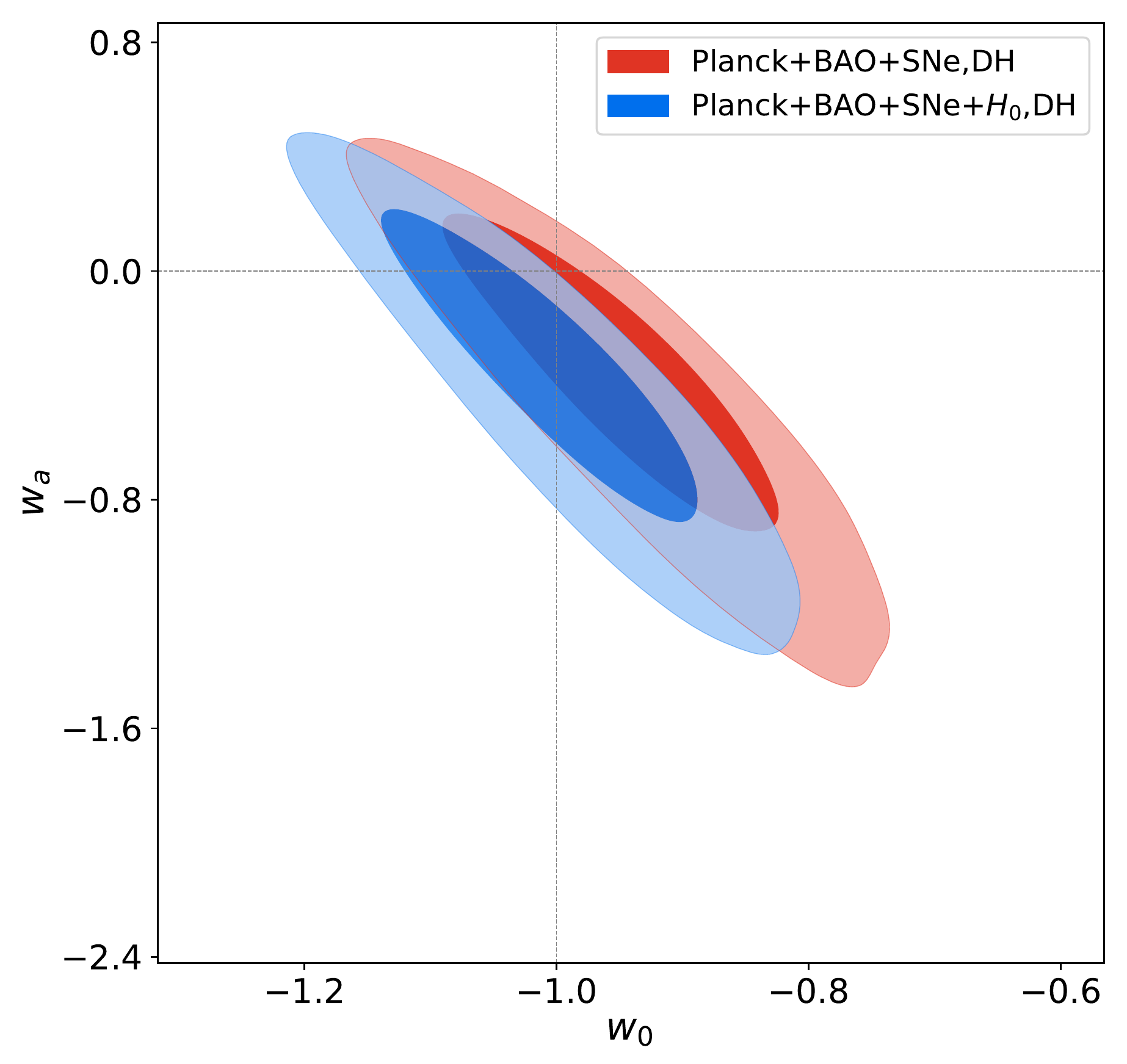}
\includegraphics[width=5.8cm]{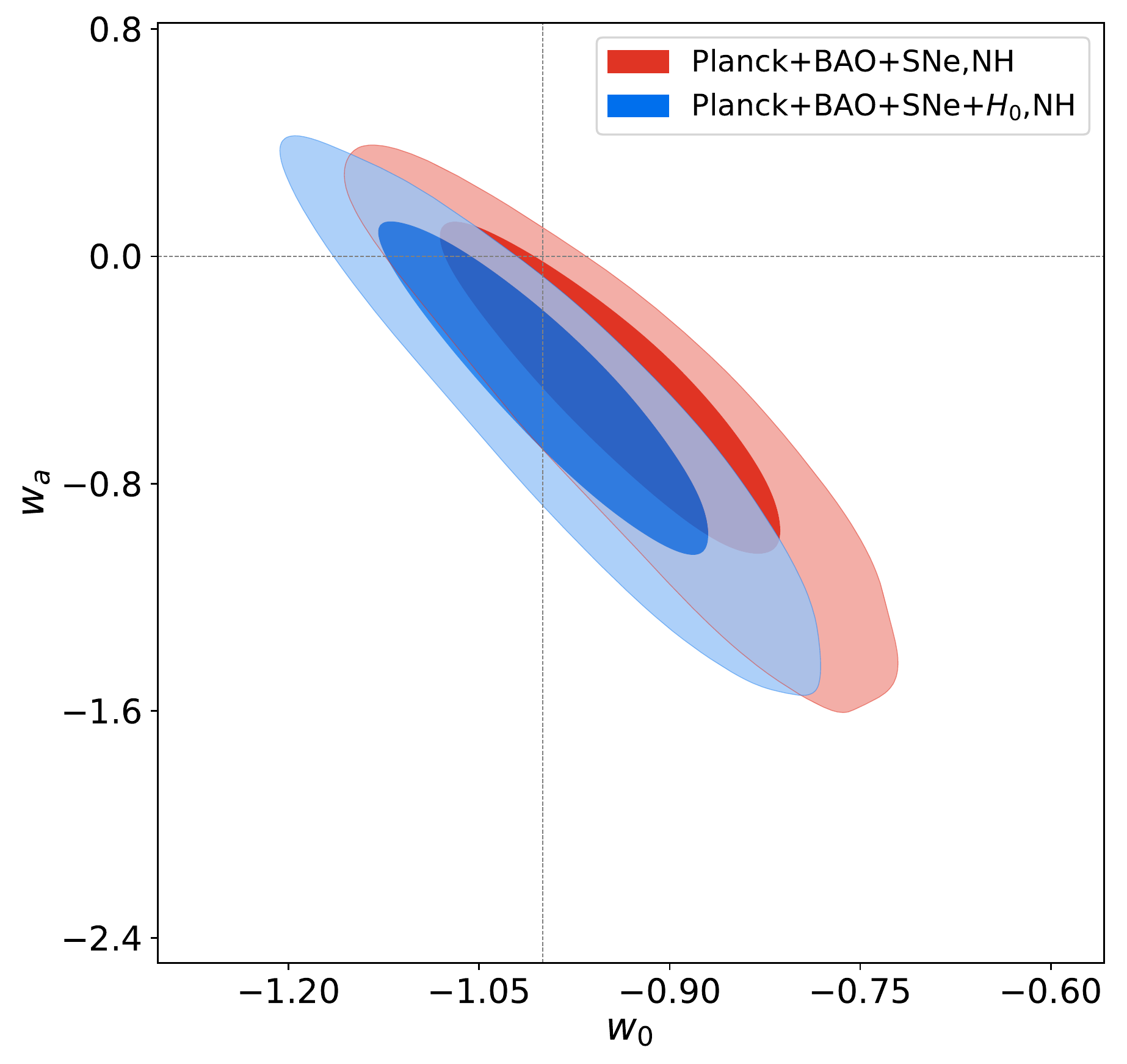}
\includegraphics[width=5.8cm]{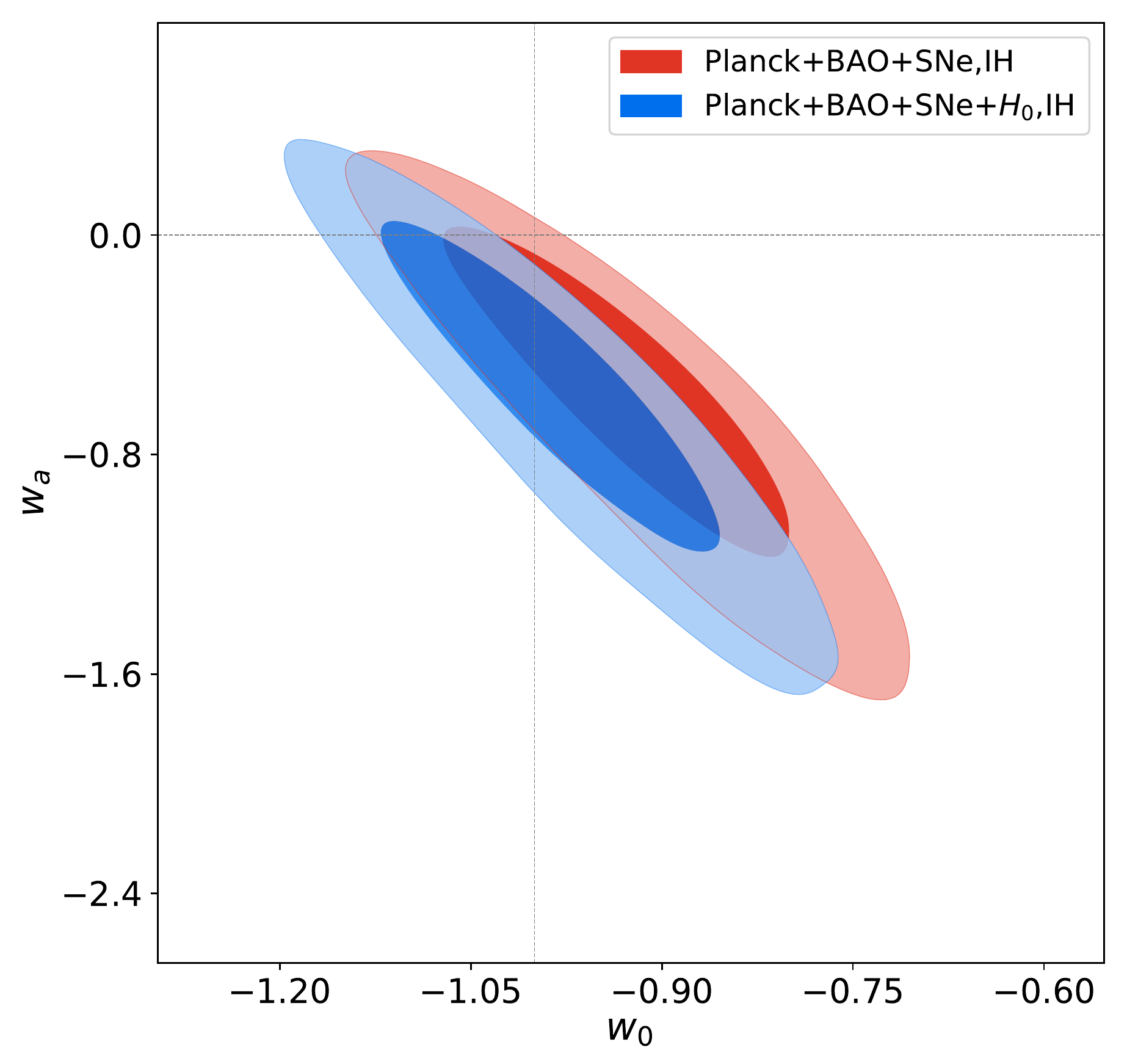}
\end{center}
\caption{The two-dimensional marginalized contours ($1 \sigma$ and $2 \sigma$) in the $w_0$-$w_a$ plane for three neutrino mass hierarchy cases, i.e., the DH case, the NH case, and the IH case, by using Planck+BAO+SNe and Planck+BAO+SNe+$H_0$ data combinations in the CPL+$\sum m_\nu$ model.}
\label{figure3}
\end{figure*}

\begin{table*}[htbp]
\caption{Fitting results of the cosmological parameters in the HDE+$\sum m_\nu$ model for three neutrino mass hierarchy cases, i.e., the DH case, the NH case, and the IH case, using the Planck+BAO+SNe and Planck+BAO+SNe+$H_0$ data combinations.}\label{table4}
\centering
\ra{1.3}
\resizebox{\textwidth}{!}{\begin{tabular}{@{}rcrrrcrrr@{}}\toprule
\hline
\multicolumn{1}{c}{Data} & \phantom{abc} & \multicolumn{3}{c}{Planck+BAO+SNe} & \phantom{abc} & \multicolumn{3}{c}{Planck+BAO+SNe+$H_0$}\\
\cmidrule{1-1} \cmidrule{3-5} \cmidrule{7-9}
\cline{1-1} \cline{3-5} \cline{7-9}
\multicolumn{1}{c}{Mass ordering} && DH & NH & IH && DH & NH & IH\\
\hline
%
%
%
%
%
$c$         && $0.645^{+0.027}_{-0.031}$  & $0.632^{+0.026}_{-0.030}$  & $0.623^{+0.025}_{-0.029}$ &
             & $0.608^{+0.023}_{-0.025}$  & $0.595\pm0.024$  & $0.587^{+0.022}_{-0.024}$ \\

$H_0$ [km s$^{-1}$ Mpc$^{-1}$]           && $67.85\pm0.81$ & $67.79\pm0.79$ & $67.74\pm0.80$ &
                                          & $69.38\pm0.72$ & $69.33\pm0.71$ & $69.27\pm0.71$ \\

$\Omega_{\rm m}$       && $0.3061\pm0.0077$ & $0.3077\pm0.0076$ & $0.3087\pm0.0076$ &
                        & $0.2927\pm0.0065$ & $0.2939\pm0.0065$ & $0.2951\pm0.0065$ \\

$\sigma_8$             && $0.797\pm0.013$ & $0.789\pm0.013$ & $0.783\pm0.013$ &
                        & $0.811\pm0.013$ & $0.803\pm0.012$ & $0.796\pm0.12$ \\

$\sum m_\nu$ [eV]      && $<0.080$  & $<0.129$ & $<0.163$ &
                        & $<0.075$  & $<0.123$ & $<0.159$   \\
\hline
$\chi^{2}$             && $3822.977$  &$3828.219$  &$3830.980$  &
                        & $3838.467$  &$3845.127$  &$3845.289$  \\
\hline
\end{tabular}}
\end{table*}

\begin{figure*}[htbp]
\begin{center}
\includegraphics[width=8.5cm]{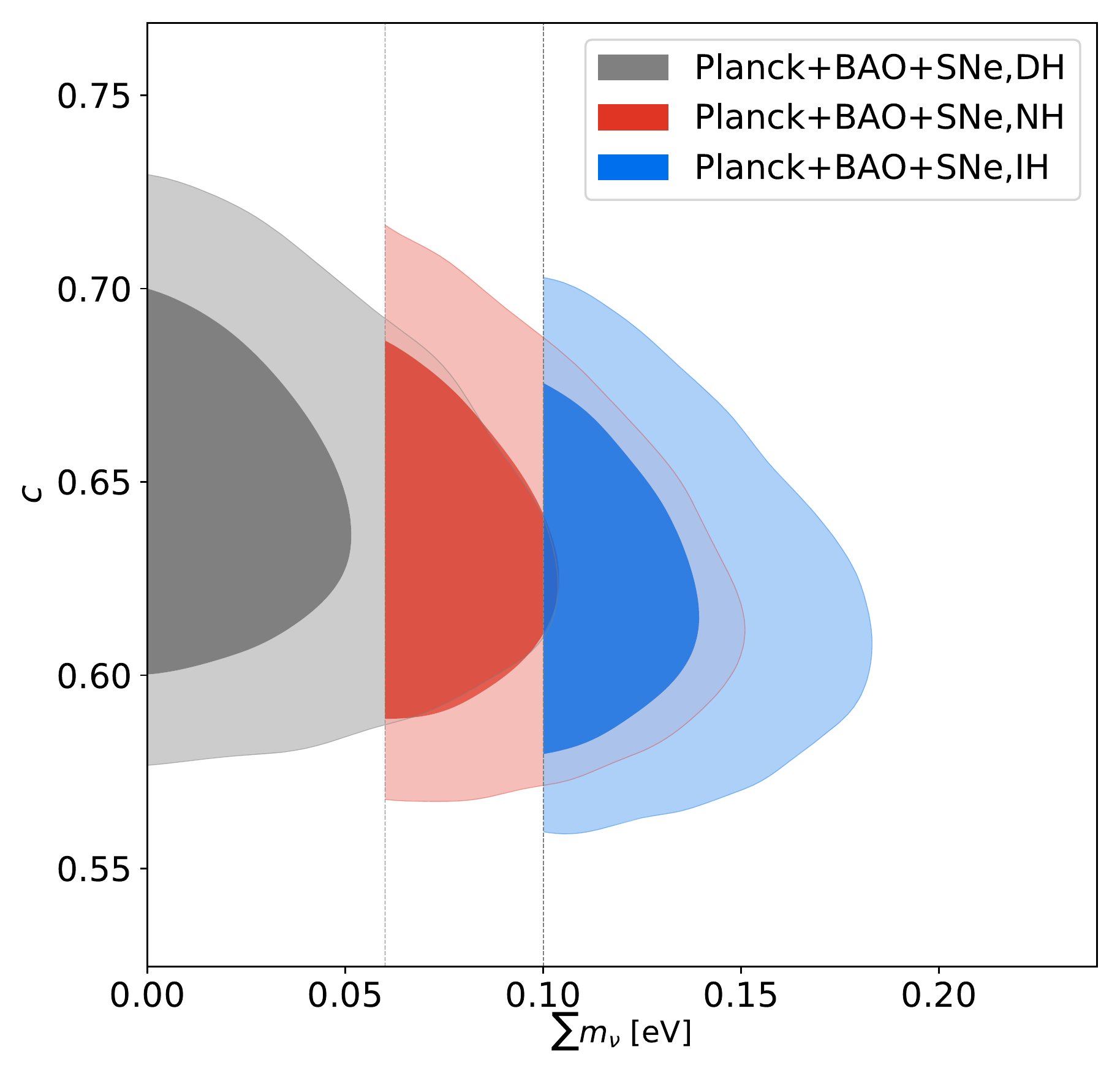}
\includegraphics[width=8.5cm]{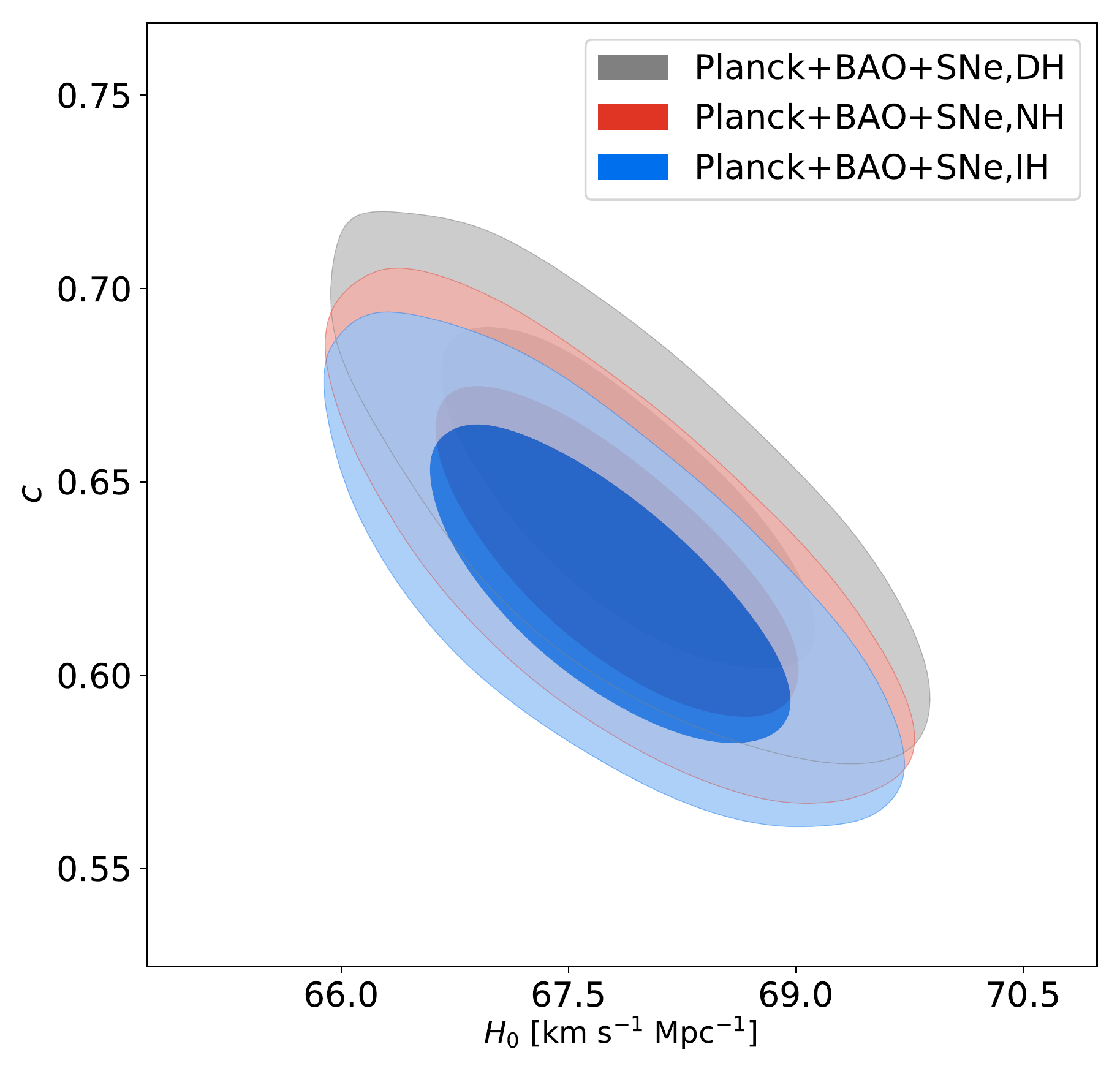}
\end{center}
\caption{The two-dimensional marginalized contours ($1 \sigma$ and $2 \sigma$) in the $H_0$-$c$ and $\sum m_\nu$-$c$ planes for three neutrino mass hierarchy cases, i.e., the DH case, the NH case, and the IH case, by using Planck+BAO+SNe data combination in the HDE+$\sum m_\nu$ model.}
\label{figure4}
\end{figure*}

In this section, we report the results of constraining the total neutrino mass from the Planck+BAO+SNe and Planck+BAO+SNe+$H_0$ data combinations. In our analysis, several typical dark energy models, i.e. the $\Lambda$CDM+$\sum m_\nu$ model, the $w$CDM+$\sum m_\nu$ model, the CPL+$\sum m_\nu$ model, and the HDE+$\sum m_\nu$ model, are investigated. In the meantime, we compare the results of the three neutrino mass hierarchy cases, i.e., the DH case, the NH case, and the IH case. The main results are listed in Tables~\ref{table1}--\ref{table4}. The best-fit values of $\chi^2$ in the various cases are also listed. The fit values of parameters are given at $68\%$ C.L. ($1 \sigma$), and the upper limits of the neutrino mass are given at $95\%$ C.L. ($2 \sigma$).

\subsection{In different dark energy models}

Firstly, we compare the constraint results in the different dark energy models from the Planck+BAO+SNe data combination. In Table~\ref{table1}, we can obtain $\sum m_\nu < 0.123$ eV for the DH case, $\sum m_\nu < 0.156$ eV for the NH case, $\sum m_\nu < 0.185$ eV for the IH case in the $\Lambda$CDM+$\sum m_\nu$ model. For the $w$CDM+$\sum m_\nu$ model, we have $\sum m_\nu<0.155$ eV (DH), $\sum m_\nu<0.195$ eV (NH), and $\sum m_\nu<0.220$ eV (IH) corresponding to $w=-1.029\pm0.035$ (DH), $w=-1.042\pm0.035$ (NH), and $w=-1.051\pm0.035$ (IH), respectively (see Table~\ref{table2}), and we find that the upper limits of $\sum m_\nu$ become larger, compared to the $\Lambda$CDM+$\sum m_\nu$ model. In the CPL+$\sum m_\nu$ model, the neutrino mass bounds are greatly relaxed (see Table~\ref{table3}), and they are $\sum m_\nu<0.247$ eV (DH), $\sum m_\nu<0.290$ eV (NH), and $\sum m_\nu<0.305$ eV (IH). As is showed in Fig~\ref{figure3}, we find that a phantom dark energy (i.e., $w<-1$) or an early phantom dark energy (i.e., the quintom evolving from $w<-1$ to $w>-1$) is slightly more favored by current cosmological observations, which leads to the fact that a larger upper limit of $\sum m_\nu$ is obtained in the $w$CDM+$\sum m_\nu$ and CPL+$\sum m_\nu$ models. For the HDE+$\sum m_\nu$ model, an early quintessence dark energy with $c<1$ (i.e., the quintom evolving from $w<-1$ to $w>-1$) is favored, and we could obtain the most stringent upper limits of the neutrino mass with $\sum m_\nu<0.080$ eV (DH), $\sum m_\nu<0.129$ eV (NH), $\sum m_\nu<0.163$ eV (IH), as also shown in Table~\ref{table4}.

In addition, we can compare the best-fit $\chi^{2}$ values of these models, which are listed in Tables~\ref{table1}--\ref{table4}. For the $w$CDM+$\sum m_\nu$ model, the $\chi^{2}$ values in the same neutrino mass hierarchy are slightly smaller than those of the $\Lambda$CDM+$\sum m_\nu$ model, at the price of adding one more parameter. We obtain the smallest $\chi^{2}$ values in the CPL+$\sum m_\nu$ model, since this model has the most free parameters. For the HDE+$\sum m_\nu$ model, the most stringent upper limits of $\sum m_\nu$ can be obtained, but the $\chi^{2}$ values are much larger than those of the $\Lambda$CDM+$\sum m_\nu$ model.

For all these models, we discuss the fitting results in the different neutrino mass hierarchies. The prior of the lower bounds of $\sum m_\nu$ are $0$ eV, $0.06$ eV and $0.1$ eV for DH, NH and IH, respectively, which can affect the constraint results of $\sum m_\nu$ significantly. In Table~\ref{table1}--\ref{table4}, the upper limits of $\sum m_\nu$ for the NH case are smaller than those for the IH case in these dark energy models. What's more, we find that the $\chi^{2}$ values in the NH case is slightly smaller than those in the IH case for all these models, which indicates that the NH case fits the current observations better than the IH case. This conclusion is still consistent with the previous studies \cite{RoyChoudhury:2019hls,Zhang:2017rbg,Guo:2018gyo,Wang:2016tsz,Feng:2019mym,Zhao:2018fjj,Zhao:2017jma,Xu:2016ddc}.

\subsection{Adding the $H_0$  measurement in data combination}

In this subsection, we report the constraint results from the Planck+BAO+SNe+$H_0$ data combination and investigate the impact of the $H_0$ measurement on the fit results of $\sum m_\nu$. As is shown in Table~\ref{table1}, we have $\sum m_\nu < 0.082$ eV for the DH case, $\sum m_\nu < 0.125$ eV for the NH case, $\sum m_\nu < 0.160$ eV for the IH case in the $\Lambda$CDM+$\sum m_\nu$ model. Adding the $H_0$ data leads to a higher $H_0$ value in the cosmological fit. From the right panel of Fig~\ref{figure1}, we can see that $\sum m_\nu$ is anti-correlated with $H_0$ in the $\Lambda$CDM+$\sum m_\nu$ model. Therefore, we obtain a smaller upper limit of $\sum m_\nu$ with the Planck+BAO+SNe+$H_0$ data combination than that with the Planck+BAO+SNe data combination, which can be clearly seen in the left panel of Fig~\ref{figure1}.

With the Planck+BAO+SNe+$H_0$ data combination, we have $\sum m_\nu < 0.145$ eV (DH), $\sum m_\nu < 0.183$ eV (NH), $\sum m_\nu < 0.210$ eV (IH) in the $w$CDM+$\sum m_\nu$ model (see Table~\ref{table2}); we have $\sum m_\nu < 0.216$ eV (DH), $\sum m_\nu < 0.255$ eV (NH), $\sum m_\nu < 0.281$ eV (IH) in the CPL+$\sum m_\nu$ model (see Table~\ref{table3}); we have $\sum m_\nu < 0.075$ eV (DH), $\sum m_\nu < 0.123$ eV (NH), $\sum m_\nu < 0.159$ eV (IH) in the HDE+$\sum m_\nu$ model (see Table~\ref{table4}). In all these models, we find that the inclusion of the $H_0$ data gives a tighter constraint on $\sum m_\nu$.

\section{Conclusion}\label{sec:4}

In this paper, using the latest cosmological observations (including the Planck 2018 CMB data), we have obtained $\sum m_\nu < 0.123$ eV (DH), $\sum m_\nu < 0.156$ eV (NH), and $\sum m_\nu < 0.185$ eV (IH) in the $\Lambda$CDM+$\sum m_\nu$ model with the Planck+BAO+SNe data combination. In addition, we also consider the influence of dynamical dark energy on the constraint results of $\sum m_\nu$. We investigate the cases of the $w$CDM+$\sum m_\nu$ model, the CPL+$\sum m_\nu$ model, and the HDE+$\sum m_\nu$ model, and we find that the nature of dark energy could significantly affect the constraints on the total neutrino mass. Compared to the $\Lambda$CDM+$\sum m_\nu$ model, the upper limits of the total neutrino mass become larger in the $w$CDM+$\sum m_\nu$ and CPL+$\sum m_\nu$ models. Using the Planck+BAO+SNe data combination, the most stringent upper limits of the neutrino mass, i.e., $\sum m_\nu<0.080$ eV (DH), $\sum m_\nu<0.129$ eV (NH), and $\sum m_\nu<0.163$ eV (IH), are obtained in the HDE+$\sum m_\nu$ model.

Comparing the values of $\chi^{2}$ between the NH and IH cases, it is found that the NH case fits the current cosmological observations better than the IH case, indicating that the neutrino mass hierarchy is more likely to be the NH case according to the current cosmological data.
In addition, it is also found that the inclusion of the local measurement of the Hubble constant in the data combination will lead to a tighter constraint on the total neutrino mass for all the dark energy model considered in this work.



\begin{acknowledgments}

We thank Hai-Li Li, Jing-Zhao Qi, and Yun-He Li for helpful discussions. This work was supported by the National Natural Science Foundation of China (Grant Nos. 11975072, 11875102, 11835009, and 11690021), the Liaoning Revitalization Talents Program (Grant No. XLYC1905011), the Fundamental Research Funds for the Central Universities (Grant No. N2005030), and the Top- Notch Young Talents Program of China.

\end{acknowledgments}

\end{document}